\documentclass[reprint,longbibliography,superscriptaddress,amsmath,amssymb,aps,prx]{revtex4-2}

\usepackage{graphicx}
\usepackage{subfigure}
\usepackage{dcolumn}
\usepackage{bm}
\usepackage{physics}
\usepackage{color}
\usepackage{float}
\usepackage{subcaption}
\usepackage[justification=raggedright,singlelinecheck=false]{caption}
\usepackage{xcolor}
\usepackage{siunitx}

\DeclareRobustCommand{\erase}{\bgroup\markoverwith{\textcolor{red}{\rule[.5ex]{2pt}{0.4pt}}}\ULon}

\begin{document}
\title{Waveguide-array-based multiplexed photonic interface for atom array}

\author{Yuya Maeda}
\affiliation{Graduate School of Engineering Science, The University of Osaka, Toyonaka, Osaka 560-8531, Japan}



\author{Toshiki Kobayashi}
\affiliation{Graduate School of Engineering Science, The University of Osaka, Toyonaka, Osaka 560-8531, Japan}
\affiliation{Center for Quantum Information and Quantum Biology, The University of Osaka, Toyonaka, Osaka 560-0043, Japan}

\author{Takuma Ueno}
\affiliation{Graduate School of Engineering Science, The University of Osaka, Toyonaka, Osaka 560-8531, Japan}

\author{Kentaro Shibata}
\affiliation{Graduate School of Engineering Science, The University of Osaka, Toyonaka, Osaka 560-8531, Japan}

\author{Shinichi Takenaka}
\affiliation{Graduate School of Engineering Science, The University of Osaka, Toyonaka, Osaka 560-8531, Japan}

\author{Kazuki Ito}
\affiliation{Graduate School of Engineering Science, The University of Osaka, Toyonaka, Osaka 560-8531, Japan}

\author{Yuma Fujiwara}
\affiliation{Graduate School of Engineering Science, The University of Osaka, Toyonaka, Osaka 560-8531, Japan}

\author{Shigehito Miki}
\affiliation{Advanced ICT Research Institute, National Institute of Information and Communications Technology~(NICT), Kobe, 651-2492, Japan} 

\author{Hirotaka Terai}
\affiliation{Advanced ICT Research Institute, National Institute of Information and Communications Technology~(NICT), Kobe, 651-2492, Japan} 

\author{Tsuyoshi Kodama}
\affiliation{Hamamatsu Photonics K.K., Shizuoka, Japan} 
\affiliation{Advanced ICT Research Institute, National Institute of Information and Communications Technology~(NICT), Kobe, 651-2492, Japan} 

\author{Hideki Shimoi}
\affiliation{Hamamatsu Photonics K.K., Shizuoka, Japan} 

\author{Rikizo Ikuta}
\affiliation{Graduate School of Engineering Science, The University of Osaka, Toyonaka, Osaka 560-8531, Japan}
\affiliation{Center for Quantum Information and Quantum Biology, The University of Osaka, Toyonaka, Osaka 560-0043, Japan}

\author{Makoto Yamashita}
\affiliation{Center for Quantum Information and Quantum Biology, The University of Osaka, Toyonaka, Osaka 560-0043, Japan}

\author{Shuta Nakajima}
\affiliation{Center for Quantum Information and Quantum Biology, The University of Osaka, Toyonaka, Osaka 560-0043, Japan}

\author{Takashi Yamamoto}
\affiliation{Graduate School of Engineering Science, The University of Osaka, Toyonaka, Osaka 560-8531, Japan}
\affiliation{Center for Quantum Information and Quantum Biology, The University of Osaka, Toyonaka, Osaka 560-0043, Japan}

\date{\today}

\begin{abstract}
The growing demand for high-capacity quantum communication and large-scale quantum computing underscores the importance of networking quantum processing units via multiplexed photonic channels. A neutral atom array with multiplexed atom-photon entanglement is a promising platform for its realization. Here, we demonstrate a key multiplexed photonic interface guiding the photons from an atom array to a single-mode waveguide array fabricated on a glass-based photonic integrated circuit. Remarkable 10 channels out of the 32-channel waveguide array with 25~$\mu \mathrm{m}$ pitch couple to photons from 10 sites of the atom array with Rydberg gate-enabled separation. Based on the observed correlation between the atomic states and the polarization of the photon with a visibility of 0.87, we anticipate its applicability to a large-scale multiplexed atom–photon entanglement generation for networking quantum processing units.
\end{abstract}

\maketitle

\section{Introduction}

Quantum networks represent a frontier in quantum information science, enabling applications such as quantum cryptography~\cite{gisin2002quantum,nadlinger2022experimental}, quantum sensing~\cite{khabiboulline2019optical,malia2022distributed}, quantum internet~\cite{kimble2008quantum,wehner2018quantum}, and distributed quantum computing (DQC)~\cite{fowler2010surface,xu2024constant,bonilla2025constant,maeda2025logical}.
Communication nodes that integrate locally computable qubits with photonic interconnections are essential for next-generation quantum networking and deploying such applications~\cite{covey2023quantum,awschalom2021development}.
Neutral atoms and ion traps are promising candidates for these communication nodes because they support high-fidelity quantum gates~\cite{paetznick2024demonstration,muniz2025high,bluvstein2025fault} and possess useful optical transitions for photonic interfaces~\cite{main2025distributed,hartung2024quantum,li2025parallelized}. While the number of qubits demonstrated on these platforms continues to expand~\cite{liu2025certified, manetsch2025tweezer}, it remains far from what is required for fault-tolerant quantum computing, motivating proposals for a large-scale DQC architectures based on ion traps~\cite{monroe2013scaling,monroe2014large}, and neutral atoms~\cite{young2022architecture,li2024high,sunami2025scalable}.
Such distributed schemes require high-rate entanglement distribution via photonic links among numerous qubits.
However, photonic entanglement generation is inherently probabilistic due to limited photon collection efficiency and transmission loss.
Therefore, improving the entanglement distribution rate through multiplexing constitutes a crucial next milestone.

Remote entanglement generation inherently involves waiting times for heralded signals, motivating temporal multiplexing schemes~\cite{hartung2024quantum,li2024high,sunami2025scalable}. Nevertheless, the achievable rate is fundamentally constrained by communication delays. Further integrating spatial parallelization improves scalability by guiding photons entangled with multiple stationary qubits to distinct photonic channels~\cite{li2025parallelized,young2022architecture,shaw2025cavity}. Such one-to-one guiding of photons requires extensive optimization of mode matching between photons and channel modes to maximize coupling efficiency and minimize crosstalk. For neutral atom array systems, so far, employing micro-optical techniques such as lensed fiber arrays~\cite{li2025parallelized} and microlens cavity arrays~\cite{shaw2025cavity}, 5-mode and 4-mode single-photon guiding to optical fibers have been demonstrated, respectively. 
\begin{figure*}[t]
    \centering
    \includegraphics[width=0.9\linewidth]{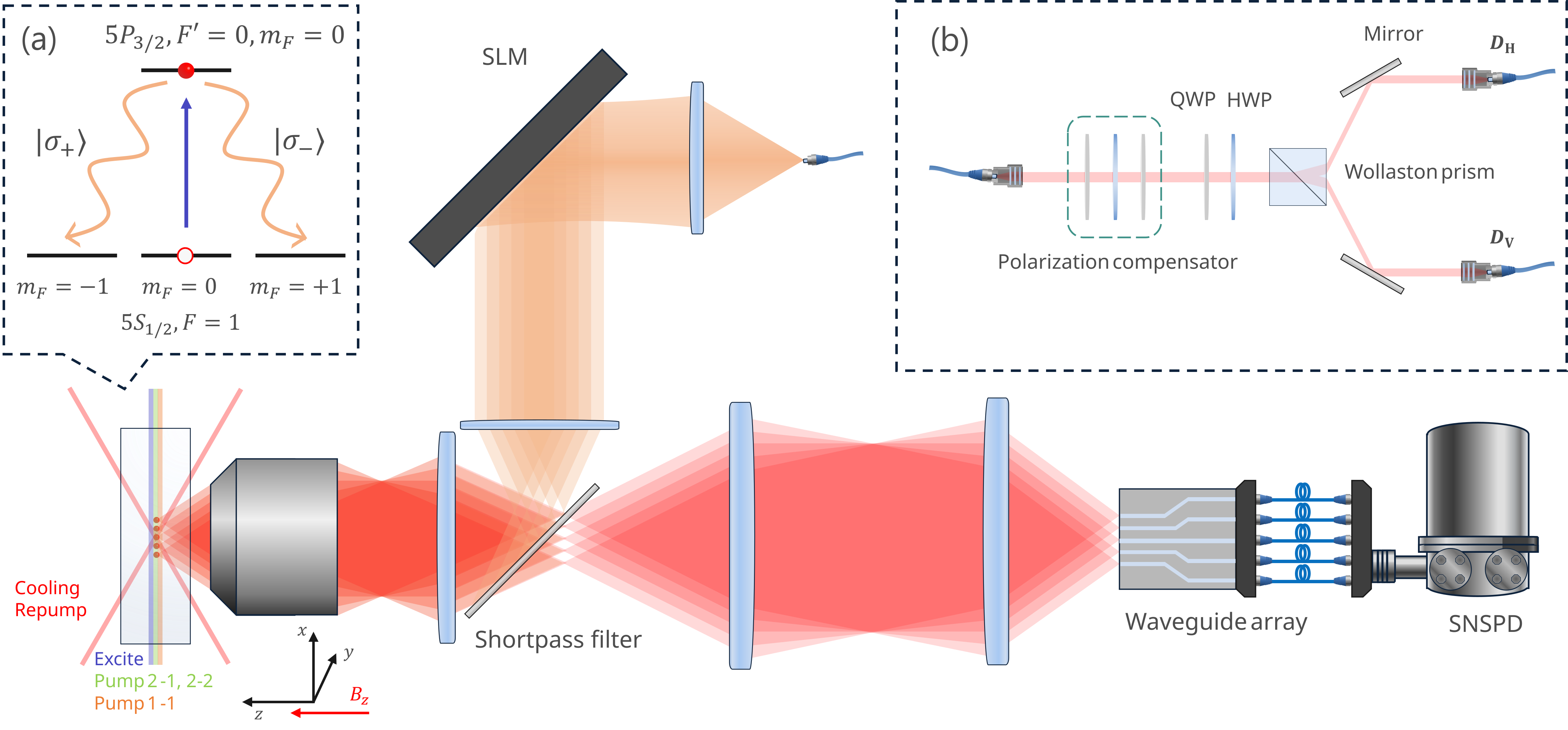}
    \caption{
    \textbf{Experimental setup.}
Our $^{87}$Rb atom array is shown by red circles at the center of the chamber. The optical tweezer array  for trapping atoms is formed by spatial light modulator~(SLM) with 852-nm laser~(drawn with orange color).
Emitted photons at 780 nm~(drawn with red color) are coupled to the waveguide array through an objective lens with $\mathrm{NA}=0.7$ and multiple lens, and then detected by multichannel superconducting nanostrip single-photon detectors~(SNSPDs). The shortpass filter passes 780-nm photons and reflects 852-nm laser for optical tweezers, thereby suppressing stray light to a negligible level. Excitation and intialization beams~(pump 2-1, 2-2 and 1-1) propagate along the $x$ axis, while the applied magnetic field defines the quantization axis along $z$ axis.
(a) Energy level diagram of atom-photon entanglement generation involving atomic Zeeman sublevels and photonic polarization states.
(b) Polarization analyzer for the collected photons. It consists of a Wollaston prism, a quarter-wave plate (QWP), and a half-wave plate (HWP). Horizontal and vertical 
polarized photons are sent to SNSPD channels $D_\mathrm{H}$ and $D_\mathrm{V}$, respectively.} 
    \label{fig:experimental_setup}
\end{figure*}

In this work, we present a scheme that achieves a higher multiplicity using a glass-based photonic integrated circuit (PIC) and report efficient coupling of photons from 10 sites of an $^{87}$Rb atom array to 10 channels from the 32-channel PIC waveguide array. Recent advances in PIC technology enable flexible and dense photonic integration. Waveguide arrays with appropriate spot sizes and spacings can be fabricated to match the input facet for an atom array and the output facet for a standard optical fiber array, enabling seamless interfacing. For example, the Teem Photonics transposer, composed of spot-size converters and waveguide-spacing concentrators, achieves efficient coupling between silicon-based PICs and fiber arrays with losses below 0.7 dB~\cite{WAFT2025waveguide}. In addition, PIC platforms can integrate components such as on-chip light sources and modulators~\cite{zhou2023prospects}, offering a compact and scalable solution for site-selective control of atom arrays.

In the present experiment, a 32-channel glass-based waveguide transposer operating at 780 nm was used to guide photons to the optical fibers. The input facet is designed to have a 25-$\mathrm{\mu m}$ spacing between nearest-neighbour waveguides. This dense waveguide array efficiently mode-matches the Rydberg-blockade-enabled 7.5-$\mathrm{\mu m}$ spacing of the atom array by using collection optics with a high numerical aperture (NA). The output facet is designed and integrated with a 32-channel optical fiber array. We optimized photon-coupling efficiencies by adjusting the atom positions using holographic optical tweezers. We measured the collection efficiencies of photons emitted from 10 single atoms into 10 corresponding waveguides, confirming negligible inter-channel crosstalk. Furthermore, we observed correlations between the atomic states encoded on Zeeman sublevels and the polarization states of the emitted photons, confirming the capability of our scheme to distribute multiplexed atom–photon entanglement.

\section{Experimental setup}
In this section, we describe our experimental setup as shown in Fig.~\ref{fig:experimental_setup}. 
Our $^{87}$Rb atom array is prepared in a glass cell of a vacuum chamber~(Infleqtion, DoubleMOT). Atomic vapor from a Rb dispenser is pre-cooled in a lower chamber with a 2D magneto-optical trap~(MOT), and then pushed up to a upper chamber with a 3D MOT.
Before trapping single atoms with optical tweezers, we cool the atoms to approximately $40~\rm{\mu K}$ by using polarization-gradient cooling~(PGC) with cooling light resonant on $F=2\leftrightarrow F'=3$ transition and repumping light resonant on $ F=1\leftrightarrow F'=1$ transition.

We trap the atoms as a one-dimensional $^{87}$Rb atom array with 10 trap sites using 852-nm holographic tweezers generated by a spatial light modulator~(SLM). The trap frequency of each traps is approximately $200~\rm{kHz}$ measured by the release and recapture method~\cite{sortais2007diffraction}.
The fluorescence photons are collected with an objective lens with numerical aperture $\mathrm{NA} = 0.7$ and coupled into a waveguide array. The waveguide array consists of 32 waveguide modes with a mode-field diameter (MFD) of $3.1~{\rm \mu m} \times 2.1~\mu$m and they are aligned at a $25~\mu$m pitch on the edge coupling side. This waveguide array is transposed to an optical fiber array of SM780 with a 127~$\mu$m pitch~(Teem Photonics, Waveguide Array to Fiber Tranposer). Further details about the waveguide array are available in Ref.~\cite{WAFT2025waveguide}. 

The photon coupling optics consist of a double 4-f system that relays and magnifies the image by a factor of 3.33; the first lens is an objective and the last lens is an aspheric lens.
This system maps the image of $7.5-\mathrm{\mu m}$ atom spacing to $25-\mathrm{\mu m}$ spacing, corresponding to the pitch of waveguide array. As illustrated in Fig.~\ref{fig:experimental_setup}, the shortpass filter reflects 852-nm trapping light and transmit 780-nm photons to separate trapping light from atomic fluorescence, effectively suppressing stray trapping light.

The collected photons are detected with superconducting nanostrip single-photon detectors~(SNSPDs).
We used 10 channels of SNSPDs out of 12 channels in a single cryostat produced by National Institute of Information and Communications Technology~(NICT) and Hamamatsu Photonics. 
The detection efficiency of SNSPD is in a range of 75-94\% and the dark count rate is less than $50~\mathrm{
s^{-1}}$.
In our estimation discussed in a later section, we assume the quantum efficiency is $\sim80\%$ as the averaged value. 

\section{Mode-match optimization between atom array and waveguide array}
\label{sec:sweep_atom_position}
\begin{figure*}
    \centering
    \includegraphics[width = \linewidth]{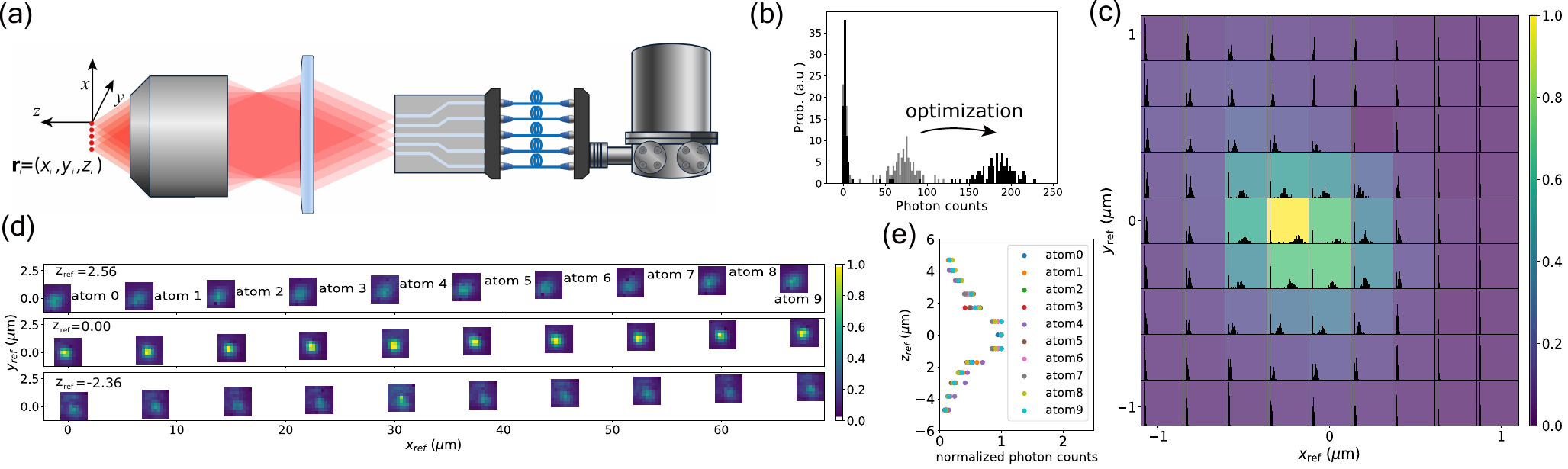}
    \caption{\textbf{Optimization of optical tweezers by photon counting experiment.} (a) Schematic of scanning the atoms by optical tweezers.
    (b) Histograms of photon counts for atom site 1 at position $r_{\mathrm{ref}}=(0.25, 0.00, 0.00)~\mathrm{\mu m}$ (gray) and $r_{\mathrm{ref}}=(-0.25, 0.00, 0.00)~\mathrm{\mu m}$ (black). The black histogram corresponds to the maximized coupling efficiency. 
    (c) 2D map of histograms obtained by scanning the optical tweezer for atom site 0 in the $xy$-plane on a $9\times9$ grid. The face color represents the total photon counts normalized by the maximum counts.
    (d) 2D maps of 10 atom sites measured at $z_\mathrm{ref} = -2.36, 0.00, 2.56~\mathrm{\mu m}$. The locations of atoms reflect actual atom spacing.
    (e) Normalized maximum photon counts obtained as a function of tweezer position along the $z$-axis.}
    \label{fig:tweezer_image}
\end{figure*}

We first introduce the method to position the atoms at the target sites achieving the optimal coupling efficiency of the emitted photons to the waveguide array by scanning optical tweezers. 
During this optimization, we shined the cooling and repumping lasers to monitor the fluorescence from the atoms. The fluorescence is detected with a 10-channel SNSPD array. Simultaneously, the target positions of the 10 atoms are scanned using the Weighted Gerchberg–Saxton (WGS) algorithm.~\cite{di2007computer,nogrette2014single}.
WGS algorithm enables us to generate holograms for the optical tweezer array that traps atoms at positions $\mathbf{r}_i = \mathbf{r}_\mathrm{ref} + i\times \Delta \mathbf{r}$, where the indices $i = 0,\ldots,9$ represent atom index and $\Delta \mathbf{r}$ denotes the spacing between adjacent atom sites.
We scanned all atom positions with varying $\mathbf{r}_\mathrm{ref}$ with fixed spacing $|\Delta \mathbf{r}|\sim7.5~\mathrm{\mu m}$, keeping relative positions of all atom.
As shown in Fig.~\ref{fig:tweezer_image}(a), we take $xy$-plane as orthogonal to the optical axis along z axis. The scale of the position is estimated from $25~\mathrm{\mu m}$ pitch of waveguide array and the lens magnification.


Figure~\ref{fig:tweezer_image}(b) shows histograms of photon counts measured on the detection channel corresponding to atom 1 at two different $\mathbf{r}_\mathrm{ref}$.
Each histogram is compiled from 200 trials consisting of a 30-ms measurement with randomly loaded atom array. These two histograms clearly show that one~(dark gray) has higher photon counts than the other~(light gray). 
We use the total photon counts in each histogram used as a metric to evaluate the coupling efficiency.

In Fig.~\ref{fig:tweezer_image}(c), we show a 2D map of the fluorescence measurement results for sweeping position $\mathbf{r}_\mathrm{ref}$ of atom site 0 over $9\times9$ grid in $xy$-plane.
The total photon counts are represented by the face color of each grid cell, and the histograms are also shown in each cell. Here we see that the optimal position for coupling to the waveguide can be found in $xy$-plane with this method.
By changing the position along with $z$-axis, we also see the defocus effect from the optimal focus position. 

In Fig.~\ref{fig:tweezer_image}(d), at three positions $z_\mathrm{ref} = 2.56, 0,$ and $-2.36~\mathrm{\mu m}$,  
we present the 2D maps of the fluorescence measurements for all the atom sites from 0 to 9. We clearly observe 
that the defocus effect appears, and the crosstalk between adjacent atom sites is sufficiently suppressed for 
the 7.5-$\mathrm{\mu m}$ atomic spacing at $z_\mathrm{ref}=0~\mathrm{\mu m}$.

Figure~\ref{fig:tweezer_image}(e) plots the maximum photon counts among all the grids in each 2D map of atom sites from 0 to 9 along with $z$ axis.
From these results, the coupling efficiency is maximized at $z_\mathrm{ref}=0.00~\mathrm{\mu m}$, where the full width at half maximum (FWHM) in the $xy$ plane and along the $z$ axis are approximately $1~\mathrm{\mu m}$ and $4~\mathrm{\mu m}$, respectively.

We note that the current optical system appears to exhibit aberrations, which are likely affecting the coupling efficiency.
The FWHM of the photon counts of $\sim 1, \sim 1$ and $\sim 4~\mathrm{\mu m}$ obtained along each of the $x, y$, and $z$ axes is larger than that predicted by an ideal, aberration-free numerical simulation~\cite{young2022architecture,robert2021thesis} of 0.85, 0.96 and 2.81, respectively.
Moreover, the measured 2D maps also suggest the presence of optical aberrations.
By correcting aberration based on the grid images obtained through the atomic position scanning, further improvement in the coupling efficiency can be expected.

\section{multiplexed single photon guiding and detection}
\label{sec:single_photon_experiment}
This waveguide array is useful for guiding multiplexed photons entangled with the atom array for quantum 
networking. In this section, we present the performance of this waveguide array with 10 multiplexed photon from 10 
atoms in atom-photon entanglement generation process~\cite{volz2005atom}. This process uses photon emission from $F'=0$ to $F=1$ transition along quantization axis as shown in Fig.~\ref{fig:experimental_setup}(a). 
The atoms are first initialized in ground state ($F=1, m_F=0$) by optical pumping with the following lasers: a 795
nm $\pi$-polarization laser resonant with the $F=1\leftrightarrow F'=1$ transition, a 780-nm $\pi$-polarization 
laser resonant with the $F=1\leftrightarrow F'=1$ transition, and a 780-nm laser resonant with 
the $F=2\leftrightarrow F'=1$ transition, having the linear polarization along x-axis.
To define quantization axis during pumping, we applied magnetic field $B\sim1.5$~G along z-axis~(see Fig.1), yielding 1.1~MHz splitting between $F=1,m_F=\pm1$ states.
The optical pumping duration is 40~$\mu\rm{s}$, resulting in a pumping efficiency of about $90\%$.

\begin{table*}[bt]
\captionsetup{justification=centering,singlelinecheck=false}
\caption{\textbf{Performance of the present waveguide-array-based photonic system} }
\label{tb:coupling_and_crosstalk}
\begin{tabular}{|l|cccccccccc|}
\hline & ch1 & ch2 & ch3 & ch4&ch5 &ch6&ch7&ch8&ch9&ch10 \\\hline
$P(p_i|a_i)\times 10^3$& $3.4$ & $3.0$ & $3.5$ & $3.9$ & $2.6$ & $1.3$ & $3.8$ & $4.7$ & $1.1$ & $1.9$ \\
$\eta_\mathrm{net}$ & $\sim0.9\%$ & $\sim0.8\%$ & $\sim0.9\%$ & $\sim1.0\%$ & $\sim0.7\%$ & $\sim0.3\%$ & $\sim1.0\%$ & $\sim1.2\%$ & $\sim0.3\%$ & $\sim0.5\%$ \\
$P(p_i|\bar{a_i})\times 10^6$ & $4.8$ & $5.3$ & $6.7$ & $4.3$ & $4.3$ & $2.4$ & $5.3$ & $7.2$ & $3.3$ & $3.3$ \\
$P(p_i|\bar{a_i})/P(p_i|a_i)$ & $0.001$ & $0.002$ & $0.002$ & $0.001$ & $0.002$ & $0.002$ & $0.001$ & $0.002$ & $0.003$ & $0.002$ \\
background & 23 Hz & 20 Hz & 48 Hz & 31 Hz & 30 Hz & 16 Hz & 27 Hz & 28 Hz & 22 Hz & 28 Hz \\\hline 
\end{tabular}
\end{table*}

After the state initialization, we shine the 100-ns pulsed excitation laser whose intensity is tuned to maximize
the spontaneous emission from the atoms~\cite{volz2006thesis}. For the characterization of the performance, we 
repeat the sequence which consists of loading atoms, the initial atom measurement, 40 trials of this photon emission process including 
state initialization and excitation, and the final atom measurement as illustrated in Fig.~\ref{fig:sequence_sketch}(a). The atom measurement ensures the presence of 
the trapped atom. This is done by the fluorescence measurement for 40-ms exposure time, described in the previous section. 
The entire sequence is repeated $1.5\times10^3$ times to achieve a total of $6.0\times10^4$ photon-generation attempts.

\begin{figure*}[t]
\centering
\includegraphics[width=\linewidth]{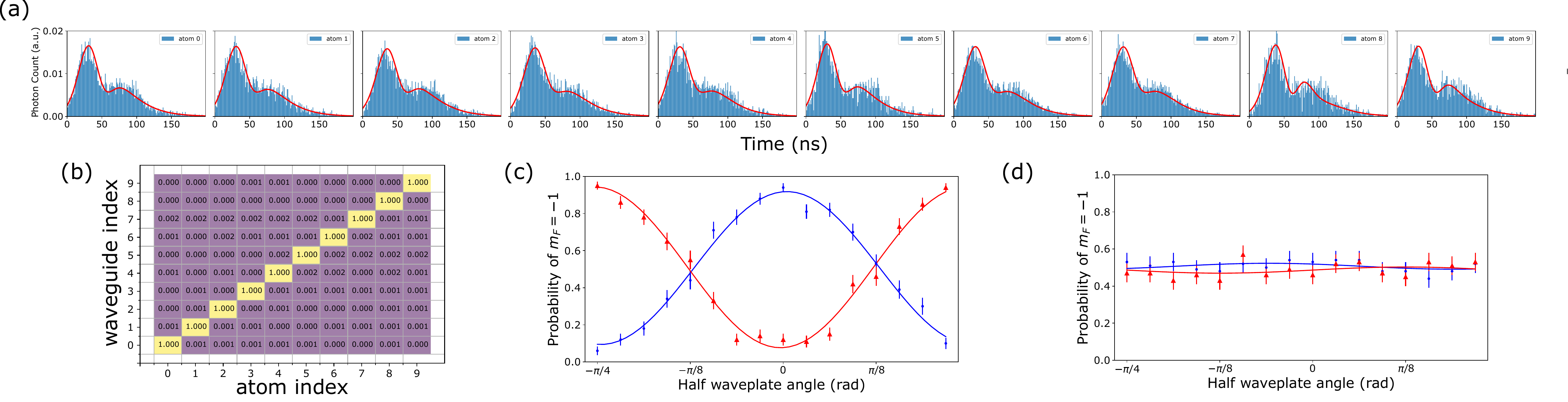}
\caption{\textbf{Characterization of  waveguide-array-coupled single photons.}~(a) Temporal profiles of spontaneous emission from atom sites 0-9. 
(b) Map of normalized photon counts from atom sites to the waveguide array. Diagonal elements represent the case where the photons emitted from atom site $i$ couple to waveguide $i$. Atom-photon correlation measurement by rotating the polarization of the photons in circular polarization basis (c) and in linear polarization basis (d).
}
\label{fig:single_photon_experiment}
\end{figure*}

In Fig.~\ref{fig:single_photon_experiment}, we present experimental results for spontaneous photon emission from the atoms. In Fig.~~\ref{fig:single_photon_experiment}(a), the normalized temporal profiles of the detected photons from atom 0 to 9 over 15000 sequence runs are shown. The normalization is performed using all photon counts within 150 ns, with each time bin of 1 ns. The red curves show the best fit to the data using the Bloch-equations-based function, with the experimentally measured excitation pulse shape and the fitting parameters such as the Rabi frequency, detuning and decoherence rate~(See Appendix.~\ref{sec:atom_excitation}). The distortion of the temporal profile is caused by the relatively slow switching of the excitation pulse, which is compared to the lifetime of D2 line. The excitation pulse is created by the acousto-optic modulator (AOM) with a 40-ns rise time. The details are discussed in Appendix~\ref{sec:atom_excitation}.

In Table \ref{tb:coupling_and_crosstalk}, we summarize the estimated performance of the present waveguide-array
based photonic system. Here, $P(p_i|a_j)$ and $P(p_i|\bar{a}_j)$ represent the photon detection probability at 
waveguide $i$ when an atom is present and absent in site $j$, respectively. In this experiment, the photon
detection time window is set to be 100 ns, and the photon counts are acquired over $1.6\times10^5$ sequence runs. 
The presence of the atom is observed with the initial atom measurement. The observed detection probabilities 
$P(p_i|a_i)$ range from 0.11 – 0.47 \%, including the efficiency of state initialization ($\sim$ 0.9), excitation 
($\eta_{\mathrm{ext}} \sim 0.67$, see Appendix~\ref{sec:atom_excitation}), optical fiber adapter ($\sim$ 0.8), and 
detection with SNSPD ($\sim$ 0.8). By excluding these contributions, the coupling efficiencies from the atoms to
the waveguides $\eta_{\mathrm{net}}$ are estimated to be 0.3–1.2 \%. These values are comparable to previous fiber
based experiments \cite{li2025parallelized,van2022entangling}. 

Furthermore, we evaluate the channel crosstalk. In Fig.~\ref{fig:single_photon_experiment}(b), we map the 
normalized photon detection ratio $P(p_i|\bar{a}_i\&a_j)/P(p_i|a_i)$, which comes from the photon detection events 
on waveguide $i$ from atom $j~(\ne i)$ when atom $i$ is absent. All measured values are below -20 dB relative to 
the diagonal elements $P(p_i|a_i)$ and are comparable to $P(p_i|\bar{a}_i)/P(p_i|a_i)$ in 
Table~\ref{tb:coupling_and_crosstalk}. In addition, $P(p_i|\bar{a}_i)=2.4-7.2~(\times10^{-6})$ in 
Table.~\ref{tb:coupling_and_crosstalk} is also comparable to the photon detection probability without excitation of 
$1.6 -4.8~(\times10^{-6})$ within the 100-ns time window, which is estimated from the background photon detection rate 
ranging from 16 to 48 Hz in Table~\ref{tb:coupling_and_crosstalk}. Note that the background photon detection rate 
is estimated by the photon detection events within 10 $\mathrm{\mu s}$ time window starting from 1 $\mathrm{\mu s}$ 
after the excitation pulse, where no photon from atoms is expected due to the observed temporal profile. These observations show that photon-detection events from atoms at other sites are 
similar to those from no atom on-site and to those from no excitation pulse. Thus, we conclude that no significant 
crosstalk arises from atoms trapped at different sites.

\section{Atom-photon correlation}
\label{sec:atom_photon_correlation}
\begin{figure}[tb]
\includegraphics[width = \linewidth]{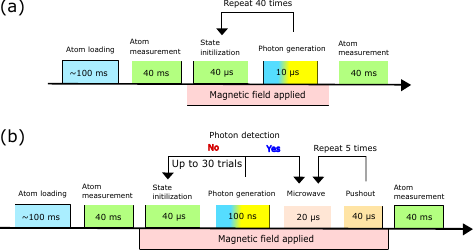}
\caption{\textbf{Experimental sequence for photon generation into waveguide array.} 
(a) Fluorescence measurement in Sec.~\ref{sec:single_photon_experiment}. The cycle includes atom loading, initial atom measurement, state initialization, excitation, and final atom measurement. The photon generation process, in which photon detection is also performed for 10$~\mathrm{\mu s}$, is repeated 40 times. (b) Atom-photon correlation  measurement in Sec.~\ref{sec:atom_photon_correlation}. The generation process is repeated up to 30 times. When the photon detection event happens, the projection measurement of the atomic state starts.
}
\label{fig:sequence_sketch}
\end{figure}

In this experimental setup, the atom and the generated photon can be in an entangled state described as 
\begin{equation}
\ket{\Phi^+} = \frac{1}{\sqrt{2}}\left(\ket{m_F=+1}_z\ket{\sigma^-}+\ket{m_F=-1}_z\ket{\sigma^+}\right),
\label{eq:bell_state}
\end{equation}
where the atomic states denoted as $\ket{m_F=\pm1}$ correspond to Zeeman sublebels of $m_F=\pm1$ at $F=1$, and the photonic states $\ket{\sigma^{\pm}}$ are circular polarization of photons.
This has been experimentally demonstrated in Ref~\cite{volz2005atom}.
In this section, we present an atom-photon correlation in this experimental setup to demonstrate the potential of 
our waveguide array system for a quantum network. As shown in Fig.~\ref{fig:experimental_setup}(b), the waveguide-
coupled photon is sent to a polarization-state analyzer and detected by SNSPDs at $D_\mathrm{H}$ and 
$D_\mathrm{V}$, where $D_\mathrm{H}$ ($D_\mathrm{V}$) detects horizontal (vertical) polarization photons. By 
properly choosing the angles of the half-wave plate (HWP) and the quarter-wave plate (QWP), we can perform the 
projection measurement of the polarization state in any basis. The projection measurement of the atomic state onto 
$\ket{m_F=-1}$ is performed as follows: we first apply a microwave $\pi$-pulse driving the 
$\ket{F=1,m_F=+1}\rightarrow\ket{F=2,m_F=+1}$ transition under a 1.5-G magnetic field. Then we push the atom in the 
state $\ket{F=2,m_F=+1}$ out of the trap site by shining a laser beam resonant with the $F=2\leftrightarrow F'=3$ 
transition. (To improve the fidelity of the projection measurement, this process was repeated five times.) After 
that, we perform the atom measurement by measuring the fluorescence from the remaining atom. In this experiment, 
after atom loading, the sequence of state initialization and photon generation is repeated up to 30 times until a
photon is detected, at which point the projection measurement of the polarization state and the atomic state is
performed, as shown in Fig.~\ref{fig:sequence_sketch}(b).

In Fig.~\ref{fig:single_photon_experiment}(c) and \ref{fig:single_photon_experiment}(d), we present the results of 
the atom-photon correlation measurement. In Fig. \ref{fig:single_photon_experiment}(c), we show the results on the 
measurement basis $\{\cos\theta\ket{\sigma^+}+\sin\theta\ket{\sigma^-}, -\sin\theta\ket{\sigma^+}+\cos\theta\ket{\sigma^-}\}$. 
The observed interference fringe exhibits typical sinusoidal behavior. Using the best-fit sinusoidal function to 
the experimental results, we estimated the visibility to be 0.82 (blue curve) and 0.87 (red curve). 
This high visibility clearly confirms the correlation between the atomic states and the photonic polarization 
states. In Fig.~\ref{fig:single_photon_experiment}(d), we show the results on the measurement basis $\{\cos\theta\ket{H}+\sin\theta\ket{V}, -\sin\theta\ket{H}+\cos\theta\ket{V}\}$. 
Here, no correlation between $\ket{m_F=-1}$ and the measurement results is observed as expected. Furthermore, for $\ket{m_F=-1}$, we measured the Bloch vector of the polarization state as $(\langle\sigma_x\rangle,\langle\sigma_y\rangle,\langle\sigma_z\rangle)=(0.09, 0.00, 0.88)$ and $(0.09, -0.01, -0.83)$ 
with detector $D_\mathrm{H}$ and $D_\mathrm{V}$, respectively. These results show that the collected photons have 
a high polarization purity of 0.89 (0.85) for $D_\mathrm{H}$ ($D_\mathrm{V}$). We also measured the Bloch vector of the polarization state without projecting the atomic state onto $\ket{m_F=-1}$. 
The results are $(\langle\sigma_x\rangle,\langle\sigma_y\rangle,\langle\sigma_z\rangle)=(0.01, -0.09, -0.02)$ and $(0.15, 0.02, 0.04)$ for $D_\mathrm{H}$ and $D_\mathrm{V}$, yielding the polarization purity of 0.50 for both cases. These results are well-explained with the state of atom-photon entanglement in Eq.~(\ref{eq:bell_state}).

\section{Discussion}
The current atom spacing of $\SI{7.5}{\mu m}$ enables a one-dimensional 200-mode multiplexing within a currently achievable $\SI{1.5}{mm}$ field of view (FOV)~\cite{manetsch2025tweezer}.
Although this spacing is larger than the range of demonstrated Rydberg gates ($ \lesssim \SI{5}{\mu m}$), the observation of Rydberg blockade has been observed using atomic ensembles separated by $\SI{10}{\mu m}$~\cite{urban2009observation}. Moreover, utilizing highly excited Rydberg states could enable high-fidelity gates at distances up to $\SI{14}{\mu m}$~\cite{saffman2025quantum}.
Therefore, the present our atom array with the waveguide-array-based photonic interface could be directly connected to a Rydberg atom-array quantum processor.

This multiplexing scheme offers various advantages depending on the application. For short-distance quantum communication applications, e.g., interconnecting quantum computers, the entanglement distribution rate scales with the number of multiplexed spatial modes. According to the communication-rate estimate for quantum error correction \cite{sinclair2025fault}, spatial multiplexing with 100 modes and an ideal free-space collection efficiency of 12\% can distribute 40 Bell pairs of atoms within 2 ms. Although the current system indicates a relatively low collection efficiency of $\sim$1\%, the aberration correction is expected to improve the efficiency. By incorporating recently developed microlens cavity-array systems \cite{shaw2025cavity}, the current system is expected to achieve more than a tenfold increase in efficiency at least. The use of cavities not only enhances photon-collection efficiency but also relaxes the mode-matching condition between the collected photons and the waveguide mode field diameter (MFD). This directly reduces the required atom spacing, substantially enhancing the feasibility of a fault-tolerant distributed quantum computation.

For long-distance quantum communication, while time multiplexing~\cite{hartung2024quantum,li2024high,sunami2025scalable} is a promising technique to increase the trial rate of entanglement generation, spatial multiplexing is also indispensable.
When a single entanglement generation attempt in the time-multiplexing protocol requires a duration $\tau~\mathrm{\mu s}$, the degree of time-multiplexing $N_{\mathrm{time}}$ for a communication distance $L~\mathrm{km}$ is limited by $N_{\mathrm{time}} \approx 5L/\tau$. Here, the term $5L$ accounts for the total waiting time, in microseconds until the success-heralding signal is received from the remote node~\cite{covey2023quantum}, assuming a light propagation speed of approximately $2\times10^8~\mathrm{m/s}$ in the fiber.
Thus, more than $N_\mathrm{time}$ qubits  at the node cannot be utilized effectively in the time multiplexing scheme only.
For example, considering the implementation of time-multiplexed entanglement generation in an atom array system \cite{manetsch2025tweezer}, sequential shuttling of atoms is necessary to couple the atoms with an optical fiber or a waveguide mode.
Assuming the atom spacing of $5~\mathrm{\mu m}$ and the shuttling-speed of $\sim0.3~\mathrm{\mu m/\mu s}$ for a single atom~\cite{bluvstein2022quantum}, the minimum duration $\tau$ determined by the shuttling time is about $20~\mathrm{\mu s}$.
When the atom array possesses $6 \times 10^3$ available qubits for communication, spatial multiplexing would be beneficial for communication distances $L$ less than $10^4~\mathrm{km}$. For 55-km communication distance, the achievable degree of time multiplexing is limited by $30$ modes. Thus, even for a long distance communication, spatial multiplexing with several hundred modes with our scheme provides substantial scalability advantages.

\section{Conclusion}
\label{sec:conclusion}
We have demonstrated a waveguide-array-based photonic interface for a neutral atom array distributed quantum computation. Installing the atom-photon entanglement generation scheme based on spontaneous emission from $F'=0$ excited state into the 10-site $^{87}$Rb atom array system, we present simultaneous 10-ch photon coupling to the 32-ch waveguide array made by glass-based PIC. The estimated coupling efficiencies are comparable to previous fiber-based experiments, and no significant crosstalk arises from atoms trapped at different sites, even though the 25-$\mathrm{\mu m}$ pitch of the present waveguides is smaller than that of standard fiber arrays. A part of the properties of atom-photon entanglement in this system, e.g., the $z$-basis correlation and the polarization states of the coupled photon, has been observed. The demonstrated multiplexed photonic interconnect technology is a key building block toward realizing networked quantum processors, enabling long-distance quantum communication and fault-tolerant distributed quantum computing.



\section*{Acknowledgement}
This work was supported by JST Moonshot R \& D Grant No. JPMJMS2066, R\&D of ICT Priority Technology (JPMI00316), ASPIRE, Grant No. JPMJAP2319, and Program for Leading Graduate Schools Interactive Materials Science Cadet Program.
YM thanks Prof. H. Weinfurter for kindly hosting a research visit and stimulating discussions.
\appendix

\section{Atom excitation}
\label{sec:atom_excitation}
Ideally, atom excitation is executed by applying a $\pi$ pulse to the $\ket{F=1,m_{F}=0}\leftrightarrow\ket{F'=0,m_F=0}$ transition after initialization. However, in practice, a strong Rabi frequency ($\Omega$) is required to suppress the effect of spontaneous emission during the pulse duration. 
The lifetime of the excited state in $^{87}$Rb is relatively short ($1/\Gamma\sim26$~ns). While this short lifetime is generally advantageous for achieving a high repetition rate due to the strong transition, it simultaneously necessitates a strong Rabi frequency ($\Omega$) and fast AOM control to effectively execute the $\pi$ pulse before spontaneous emission occurs~($\Omega\gg\Gamma$).
We simulate the excitation scheme using the two-level density matrix model defined in Eq.~\ref{eq:bloch}.
Let $\rho$ be the density matrix for the two-level system involving $\ket{F=1,m_F=0}$ and $\ket{F'=0,m_F=0}$, and let $\Gamma$ be the decay rate of the D2 line. Due to the branching ratio, one-third of the spontaneous emission returns to the initial state $\ket{F=1, m_F=0}$, while the remaining two-thirds result in the emission of $\sigma^+$ or $\sigma^-$ polarized photons, which are measured as the signal.
The resonant Rabi frequency, the detuning of the excitation pulse, and the dephasing rate are denoted by $\Omega$, $\delta$, and $\gamma$, respectively.
\begin{equation}
\label{eq:bloch}
\begin{aligned}
\dot{\rho}_{11} &= -\frac{i\Omega}{2}(\rho_{12} - \rho_{21})+\frac{\Gamma}{3}\rho_{22} \\
\dot{\rho}_{22} &= \frac{i\Omega}{2}(\rho_{12} - \rho_{21}) - \Gamma \rho_{22} \\
\dot{\rho}_{12} &=(-i\delta- \gamma)\rho_{12} + \frac{i\Omega}{2}(\rho_{22} - \rho_{11}) \\
\dot{\rho}_{21} &= (i\delta-\gamma)\rho_{21} - \frac{i\Omega}{2}(\rho_{22} - \rho_{11})
\end{aligned}
\end{equation}
The total strength of emitted photons is proportional to the population of excited state in the considered two levels. We calculate the population to estimate the excitation pulse parameters by optimizing the fit to the obtained photon counts. The optimized parameters include the timing, width, peak intensity, detuning, of excitation pulse and dephasing rate between $\ket{F=1,m_F=0}$ and $\ket{F'=0,m_F=0}$. We normalize the population to fit the obtained photon counts, accounting for experimental factors such as finite coupling efficiency.
The result of the fitting, as shown in Fig.~\ref{fig:single_photon_experiment}(a), is consistent with our experimental data. Specifically, we obtained an excitation-pulse FWHM of 
$23.26\pm0.71$~ns, which is consistent with the FWHM of 23.8 ns measured with a photodetector. Furthermore, the fitted peak intensity corresponds to a Rabi frequency of $\sim21.55\pm1.8$ MHz, explaining the gentle slope observed in our spontaneous emission result. We also obtained a dephasing rate of $2.87\pm0.23~\rm{MHz}$ and a detuning of $-0.38\pm0.35~\rm{MHz}$.
From these estimations, we were able to estimate the excitation efficiency as $\eta_{ext}\sim0.67$ by calculating the leakage of population $(1-\rho_{11}-\rho_{22})$ from considered two levels contributing to detectable photon emission during the time window of the SNSPDs.
The obtained signal includes the second cycle of spontaneous emission, as referred to in Sec.\ref{sec:single_photon_experiment}, but this does not influence the visibility result since a $\pi$-polarized photon is not coupled into the waveguide along the quantization axis~\cite{hofmann2013thesis}.


\section{Error analysis}
\label{sec:error_analysis}

Figure~\ref{fig:single_photon_experiment}(c) shows survival probability of an atom obtained by the sequence described in Sec.~\ref{sec:single_photon_experiment}, which illustrates correlation between polarization of detected photons and atomic spin states.
The fitting function to give the visibility is expressed as
\begin{equation}    
    f(\theta)=A\sin\left(\theta+B\right)+C,
    \label{eq:sine_fitting}
\end{equation}
where $A,B$ and $C$ are fitting parameters~\cite{volz2005atom}, and $\theta$ is a rotated angle of HWP converted to radian. 
In the ideal case, both the amplitude $A$ and the offset $C$ are $1/2$. However, we obtained $A=0.41~(-0.43)$ and $C=0.51~(0.51)$ with respect to photon detection events at the detector $D_\mathrm{H}$ and $D_\mathrm{V}$.

The dominant factor limiting visibility might be imperfections in the measurement of the atomic states. We observed a heating effect caused by the repeated trials of state initialization and excitation pulses which led to an atom loss of approximately $5\%$. The beam used to push atoms in the $F=2$ state out of the traps also introduces a few percent error due to off-resonant transition to $F'=2$. 

Another possible error source is the mismatch between the optical axis defined by waveguide mode and the quantization axis of atoms defined by applied magnetic field. The mismatch can mix undesired components to the polarization of photons. In this case, the transition shown in Fig.~\ref{fig:experimental_setup}(c) produces elliptically polarized photons in the waveguide mode.
The magnitude of this effect is not directly measured. However, if the waveguide mode is tilted by an angle $\theta$ to the quantization axis, the ellipticity of generated photons $\chi$ satisfies $\tan\chi=\cos\theta$, which gives an ellipticity of 0.98 when $\theta=0.17~\rm{rad}$. 

On the other hand, the influence of background noise-such as stray photons and dark counts of SNSPD are negligible error source in current visibility regime. The SNR for each port of the polarization analyzer is more than $5\times10^2$, so the degradation of visibility could be less than $0.2~\%$.

The error arising from misalignment in the polarization compensation between the single-mode fiber and the polarization analyzer is also expected to be small.
We connected the waveguide module and the polarization analyzer with a single-mode fiber, which caused the polarization of the coupled photons to rotate.
We compensated for the rotation, as well as the difference in the polarization reference axes between the waveguide module on the optical table and the polarization analyzer, using a polarimeter.
To implement this polarization compensation, we propagated light backward from the polarization analyzer and monitored its polarization state with a polarimeter while adjusting two QWPs and one HWP.
From the obtained stokes parameters, the residual polarization error can be kept below 0.1\%, provided that the polarimeter is not tilted to the waveguide mode. 
We also observed that the polarization drifts by one percent over the course of a day, indicating that the polarization rotation induced by single-mode fiber is not the dominant error source.

\bibliography{refs.bib}
\end{document}